\documentstyle[epsfig,12pt]{elsart} 
\begin{document}
\newcommand{\kp}{K^+}
\newcommand{\gk}{\vec{\gamma}\vec{k}}
\newcommand{\gE}{\gamma_0 E_k}
\newcommand{\ppl}{\vec{p}}
\newcommand{\bcm}{\vec{b}^{\star}}
\newcommand{\becm}{\vec{\beta}^{\star}}
\newcommand{\bepl}{\vec{\beta}}
\newcommand{\rcm}{\vec{r}^{\star}}
\newcommand{\rpl}{\vec{r}}
\newcommand{\A}{{$\mathcal A$}}
\newcommand{\wpk}{ \omega_{p-k}}
\newcommand{\Journal}[4]{ #1 {\bf #2} (#4) #3}
\newcommand{\NPA}{Nucl.\ Phys.\ A}
\newcommand{\PLB}{Phys.\ Lett.\ B}
\newcommand{\PRC}{Phys.\ Rev.\ C}
\newcommand{\ZPC}{Z.\ Phys.\ C}
\newcommand{\be}{\begin{equation}}
\newcommand{\ee}{\end{equation}}
\begin{frontmatter}

\title{Formation of double-$\Lambda$ hypernuclei at PANDA}

\author{T.~Gaitanos,}
\author{A.B.~Larionov$^{*}$,}
\author{H.~Lenske,}
\author{U.~Mosel}

\address{Institut f\"ur Theoretische Physik,
Universit\"at Giessen, D-35392, Giessen, Germany}
\address{${}^{*}$Also at: Russian Research Center, 
Kurchatov Institute, 123182 Moskow, Russia}
\address{email: Theodoros.Gaitanos@theo.physik.uni-giessen.de}
\begin{abstract}
We study the formation of single- and 
double-$\Lambda$ hypernuclei in antiproton-induced reactions relevant 
for the forthcoming PANDA experiment at FAIR. 
We use the Giessen Boltzmann-Uehling-Uhlenbeck (GiBUU) 
transport model with relativistic mean-fields for the description of 
non-equilibrium dynamics and the statistical multifragmentation model 
(SMM) for fragment formation. 
This combined approach describes the dynamical properties of strangeness 
and fragments in low energy $\bar{p}$-induced reactions fairly well. 
We then focus on the formation of double-$\Lambda$ hypernuclei in 
high energy $\bar{p}$-nucleus collisions on a primary target including the 
complementary $\Xi$-induced reactions to a secondary one, as proposed by the 
PANDA Collaboration. Our results show that a copious production of 
double-$\Lambda$ hyperfragments is possible at PANDA. In particular, we 
provide first theoretical estimations on the double-$\Lambda$ 
production cross section, which strongly rises with decreasing energy 
of the secondary $\Xi$-beam. 
\end{abstract}
\begin{keyword}
BUU transport equation, Giessen-BUU, statistical multifragmentation model, 
antiptoron-induced reactions, hypernuclei, 
PANDA experiment.\\
PACS numbers: {\bf 25.75.-q}, {\bf 21.65.+f}, 21.30.Fe, 25.75.Dw. 
\end{keyword}
\end{frontmatter}

\date{\today}

\section{Introduction}
\label{sec1}

The investigation of hypernuclei is directly related to 
various fundamental aspects of nuclear and hadron 
physics~\cite{Haidenbauer:2007ra,friedmann}. Of extreme interest 
are still incompletely known hyperon-nucleon and hyperon-hyperon 
interactions. 
The knowledge of these interactions is crucial for a better 
understanding of the strangeness sector of the hadronic equation 
of state, in particular, beyond ground state baryon density. 
For instance, compact astrophysical objects such 
as neutron stars might be strongly affected by the presence of hypermatter 
due to a considerable softening of the hadronic equation of state at 
very high baryon 
densities~\cite{Glendenning:1997wn,SchaffnerBielich:2010am,Hofmann:2000mc}, 
which, however, seems to be in contrast with the recently observed neutron star 
of about two solar masses~\cite{demorest}. 
Hypernuclear structure provides information around 
saturation density only, while the production of hypermatter in nuclear 
reactions induced by heavy ions and by antibaryons covers a broader  
region in baryon density, isospin asymmetry and single-particle 
energy~\cite{Saito:2010zz,Hashimoto:2010zza,SanchezLorente:2008zz,Pochodzalla:2005nf}. 
Nuclear reactions further can be useful for the formation of exotic hypernuclei 
and help to understand the limits of the nuclear chart also into the strangeness 
sector~\cite{KEK}. Furthermore, the investigation 
of hypernuclei is important for spectroscopy of conventional nuclei 
\cite{Hiyama:2010zzb,Hiyama:2010zz,Lorente:2011ev}, since single hyperons
bound in finite nuclei do not experience the Pauli blocking and 
thus serve as probes for many-body dynamics. 

The hypernucleus is a bound nucleus with one or more nucleons replaced 
by $\Lambda$-hyperons. So far only single-$\Lambda$ (or $S=-1$ single-strange) and 
double-$\Lambda$ (or $S=-2$ double-strange) hypernuclei have been found experimentally. 
In the first case the 
physical properties of the hyperon-nucleon interaction can be studied, 
while in double-$\Lambda$ hypernuclei also the hyperon-hyperon interaction 
is accessible.

Experimental information on single-$\Lambda$ hypernuclei is conventionally 
provided by spectroscopy using pion or kaon beams, by high energy protons, 
and by electroproduction~\cite{hypprod1,hypprod1b}. In these cases 
the structure of rather cold hypernuclei at ground state 
density is explored. In reactions induced by intermediate energy 
heavy-ion beams, however, a quite different scenario is encountered: 
hyperons are produced at densities higher 
than saturation, and then can be captured by nuclear fragments. 
Therefore, in such reactions one might explore 
indirectly the high density behaviour of the hyperon-nucleon interaction. 
The production of single-$\Lambda$ hypernuclei 
in reactions between heavy nuclei was first theoretically proposed by 
Kerman and Weiss~\cite{kerman}. Complementary studies 
then followed by several groups~\cite{rudy,others1,wakai}. 
Very recently experiments at JLAB on 
hypernuclear spectroscopy have been started~\cite{Hashimoto:2010zza}. 
Recent observations on hypernuclei and antihypernuclei in relativistic 
heavy-ion collisions have been reported recently by the STAR 
Collaboration~\cite{STAR} (see also for an experimental overview 
Ref.~\cite{PochoExp}). Furthermore, the 
HypHI and FOPI Collaborations~\cite{hyphi1,hyphi2,fopihyp} 
at GSI have performed heavy-ion experiments where the experimental analysis 
on hypernuclei are still under progress. Also recent theoretical 
investigations have been started~\cite{gaitanos1}. 

The formation of double-$\Lambda$ hypernuclei is in the focus of 
strangeness-nuclear physics since the experimental discovery of the 
${}^{10}_{\Lambda\Lambda}$Be~\cite{Danycz63} and ${}^{6}_{\Lambda\Lambda}$He 
~\cite{Pnowse66} hypernuclei in the 60's by measuring their double pion decay 
(see Refs.~\cite{Pochodzalla:2005nf,KEK2} for the present experimental status of 
$\Lambda\Lambda$-hypernuclei). For this purpose the copious 
production of rather slow $\Xi$-hyperons is necessary. One of the key projects 
in the new FAIR facility is the experimental investigation of double-strange 
hypernuclei by the PANDA Collaboration~\cite{panda1,panda2,panda3,ferro}, 
which is the main subject of this paper.  
Here one intends to form hypernuclei by reactions induced by antiprotons ($\bar{p}$) at 
beam momenta around $3$ GeV/c, i.e., close to the $\Xi\bar{\Xi}$ production 
threshold ($P_{\rm lab}^{\rm thr}=2.62$ GeV/c). 
In contrast to heavy-ion collisions and proton-induced reactions, where strangeness 
production proceeds mainly through meson rescattering and resonance decay, 
here the main production mechanism for hypernuclei arises from 
$\rm \bar{p}p$- and $\rm \bar{p}n$-annihilation. This channel has very high 
cross section at PANDA energies, i.e., $\sigma_{\rm \bar{p}p}\simeq (60-80)$ mb, 
which is almost 
a factor of $2$ higher than the corresponding $pp$ cross section~\cite{PDG}. 
According to the PANDA proposals, double-strange hypernuclei are expected to 
be produced through the capture of primary cascade particles ($\Xi$) into secondary 
targets, which then are converted inside a nuclear fragment into two $\Lambda$ 
hyperons. 

Recently we have studied the formation of S=-1 single-strange hypernuclei in 
heavy-ion collisions and high energy proton-induced reactions~\cite{gaitanos1}. 
Our theoretical estimates on the production probability of strangeness 
carrying nuclei seem to be compatible with the HypHI experiment~\cite{hyphi3} 
and are also close to other theoretical analyses~\cite{wakai}. 
Here we extend our previous 
investigations to the formation of double-strange hypernuclei in antiproton 
induced reactions at energies close to those of the proposed PANDA experiment~\cite{panda2}  
at FAIR. The 
theoretical treatment is based on a relativistic kinetic theory in the 
framework of the GiBUU transport model \cite{gibuu} supplemented by a statistical 
model of fragment formation~\cite{smm}. The theoretical background is 
presented in Section \ref{sec2}. Detailed results on antiproton-induced 
reactions are then presented and discussed in Section \ref{sec3}. Conclusions 
and final remarks close this work in Section \ref{sec4}.

\section{Transport theoretical framework}
\label{sec2}

The fast non-equilibrium stage of hadron-nucleus reactions is described 
by the GiBUU 
transport model. GiBUU consists of a common transport 
theoretical framework for various reaction types, e.g., 
heavy-ion collisions, photon-, electron-, neutrino- and hadron-induced 
reactions. For a detailed description see~\cite{gibuu}. Here we apply 
it to antiproton-induced reactions at PANDA energies and to reactions 
induced by low-energetic $\Xi$-hyperons.

The GiBUU model is based on a covariant extension of the semiclassical 
kinetic theory~\cite{kada,botermans}, the relativistic 
Boltzmann-Uehling-Uhlenbeck equation, which reads
\begin{equation}
\left[
k^{*\mu} \partial_{\mu}^{x} + \left( k^{*}_{\nu} F^{\mu\nu}
+ m^{*} \partial_{x}^{\mu} m^{*}  \right)
\partial_{\mu}^{k^{*}}
\right] f(x,k^{*}) = {\cal I}_{coll}
\quad .
\label{rbuu}
\end{equation}
Eq.~(\ref{rbuu}) describes the dynamical evolution of the one-body phase-space 
distribution function $f(x,k^{*})$ for the hadrons under the influence of a 
hadronic mean-field (l.h.s. of Eq.~(\ref{rbuu})) and binary collisions. 
In the spirit of the 
relativistic mean-field (RMF) approximation of Quantumhadrodynamics the kinetic 
equation is formulated in terms of kinetic $4$-momenta 
$k^{*\mu}=k^{\mu}-\Sigma^{\mu}$ and effective (Dirac) masses $m^{*}=M-\Sigma_{s}$. 
The self-energy is given by its Lorentz-vector ($\Sigma^{\mu}$) and 
Lorentz-scalar ($\Sigma_s$) components
\begin{eqnarray}
\Sigma^{\mu} & = & g_{\omega}\omega^{\mu} + \tau_{3}g_{\rho}\rho_{3}^{\mu}
\nonumber\\
\Sigma_{s} & = & g_{\sigma}\sigma
\quad ,
\label{SelfEnergies} 
\end{eqnarray} 
with the isoscalar, scalar $\sigma$, the isoscalar, vector $\omega^{\mu}$ meson fields. 
$\rho_{3}^{\mu}$ is the third isospin-component of the isovector, vector 
meson field, and $\tau_{3}=\pm 1$ for protons and neutrons, respectively. 
The self-energies describe the interaction between nucleons inside infinite 
nuclear matter, where the classical meson fields obey the standard Lagrangian equations 
of motion~\cite{qhd}. The meson-nucleon coupling constants $g_{i}$ 
($i=\sigma,\omega,\rho$) and also the additional parameters of the non-linear 
self-interactions of the $\sigma$ meson (not shown here) are taken from the 
widely used NL parametrizations~\cite{lala}. 

Since we will study antiproton-nucleus reactions, we need the antinucleon-meson 
vertices, too. In principle, the description of 
the in-medium antinucleon interaction 
should be based on the properties of the strong interaction  under 
G-parity transformation. G-parity transformation of the nucleon field, as 
in Eq.~(\ref{SelfEnergies}) is 
equivalent with a sign change of only the isoscalar Lorentz-vector component of the 
baryon self-energy. However, the conventional RMF approach fails to describe 
empirical data on antiproton production in $p$-nucleus and nucleus-nucleus 
interactions by imposing only G-parity 
\cite{cassing}. This problem has been related to a linear energy dependence 
of the proton-nucleus optical potential, which is not consistent with empirical data 
and results to a divergent linear behavior for the antinucleon-nucleus interactions. 
A way to overcome this issue has been recently proposed in~\cite{nld1,nld2,nld3}, 
using non-linear derivative interactions. Indeed, not only the energy dependence of 
the real part of the in-medium antinucleon interaction turned out to be 
in agreement with available data~\cite{nld2}. 
Also the imaginary (dispersive) part of the in-medium antinucleon 
optical potential significantly contributes to the energy dependent 
in-medium antinucleon interaction, again in agreement with empirical studies~\cite{nld2}. 
For the description of complex nuclear dynamics such as 
heavy-ion collisions within transport theory a simpler approach for the RMF 
fields is applied here. The antinucleon-meson coupling constants are rescaled by 
a phenomenological parameter such, to reproduce experimental data in 
high energy antiproton-induced reactions~\cite{larionov}. 

Numerically, the collision term is treated in a parallel ensemble 
algorithm incorporating the standard parametrizations for the cross sections 
of various binary processes~\cite{gibuu}. Important for this work is the recent 
implementation in GiBUU~\cite{larionov,larionov2}
of $\bar{N}N$-annihilation channels according to a statistical 
model of Pshenicknov et al.~\cite{StatAnn}, including $\bar{K}KX$ mesonic 
final states. 
Furthermore, the following primary channels have been 
implemented recently ($Y,N$ stand for hyperons and nucleons, respectively):
$\bar{N}N\rightarrow \bar{Y}Y$, where the channel 
$\bar{N}N\rightarrow \bar{\Xi}\Xi$ is of particular interest. 
Their parametrizations are given in Refs.~\cite{larionov2,faessler}. 
Secondary interactions involving hyperons contribute also to the production 
of hypernuclei. Among the various secondary 
processes, the 
channels $\bar{K}N\rightarrow \Xi K$ and $\Xi N\rightarrow \Lambda\Lambda$ are 
essential for hypernuclear formation. The cross section of the former binary 
collision is taken from~\cite{hypnuc1,hypnuc1a,hypnuc2,baldini}, 
while the cross section for the latter process is taken from a parametrization 
to theoretical calculations~\cite{esc04}.

The present implementation of the GiBUU model turned out to work well for 
strangeness production at PANDA energies, as shown in Ref.~\cite{larionov2}. 
However, for the description of single- and double-$\Lambda$ hypernuclei 
not only the yields and spectra of produced strange particles 
should be in agreement with experiment, but also the fragmentation 
of the residual target source. As a well-known problem, Boltzmann-like 
transport equations do not provide information of the dynamical evolution 
of physical fluctuations. Thus, one has to model fragment formation by 
coalescence or using complementary statistical models. 
Previously, we have applied the statistical multifragmentation model (SMM) from 
Botvina et al.~\cite{smm} to spectator fragmentation and to the decay of 
residual nuclei in reactions induced by protons and heavy 
ions~\cite{gaitanos1,gaitanos2} for the formation 
of nuclear fragments. It turned out that the hybrid GiBUU+SMM 
approach reproduces the data on fragmentation with high accuracy. 
The combination between the pre-equilibrium dynamics (GiBUU) and the 
fragmentation stage (SMM) in hadron-induced reactions consists of the following 
steps: at first, we calculate the average mass 
$\langle A \rangle$, charge $\langle Z \rangle$ and excitation 
energy $E^*$ of the bound system (residual source) as a function of time, 
where bound particles are selected by a simple density cut. The excitation 
energy is obtained event-by-event from energy balance. When the average properties 
of the residual bound system do not experience any time dependencies, we 
stop the simulation and switch to the SMM model. We also determine the 
anisotropy ratio at the central shell of the residual source. It 
gives complementary information on the degree of local equilibration 
of the system, which is required by the SMM model. 

\begin{figure}[t]
\unitlength1cm
\begin{picture}(10.,5.0)
\put(1.75,0.){\makebox{\psfig{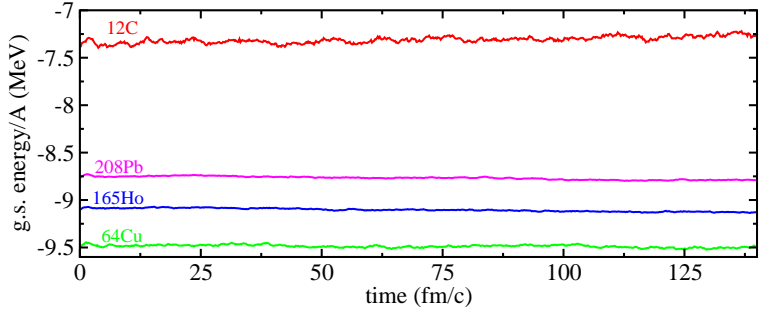}}}
\end{picture}
\caption{Ground state (g.s.) energy per nucleon (subtracting the 
nucleon mass) as function of time for different targets, as indicated. 
The calculations were performed in the Vlasov mode, i.e., without 
including binary collisions.
}
\label{Fig1}
\end{figure}
Numerically the GiBUU equations are solved within the test particle method, 
where the phase-space distribution function is discretized by so-called 
test particles~\cite{testpart}. One then solves the relativistic Hamiltonian 
equations of motion for the test particles, while the collision term is 
simulated by standard Monte Carlo prescriptions. In such computational 
simulations spurious numerical noise is unavoidable. 
Important for the present study is 
a very good stability of the ground state nucleus, which is not obvious 
within the test particle formalism. A very good nuclear ground state stability 
is also required, in order to avoid spurious particle 
emission and non-conservation of energy. This issue is important when 
determining the excitation energy of a residual system. As described in 
detail in Ref.~\cite{rtf}, we use the same energy density functional 
to extract the ground state density profiles needed for the initialization 
and to propagate then the system. The results of this improved 
initialization prescription are shown in Fig.~\ref{Fig1}. As one can see, 
the ground state energy is almost perfectly stable in time. In particular, for the 
heavier systems energy conservation is perfect. More details on this study 
can be found in Ref.~\cite{rtf}. 
We therefore expect that the excitation 
energy in reactions with these nuclear targets is extracted in a reliable way 
by reducing spurious contributions as much as possible.

\section{Results}
\label{sec3}

\begin{figure}[t]
\unitlength1cm
\begin{picture}(10.,8.0)
\put(1.75,0.){\makebox{\psfig{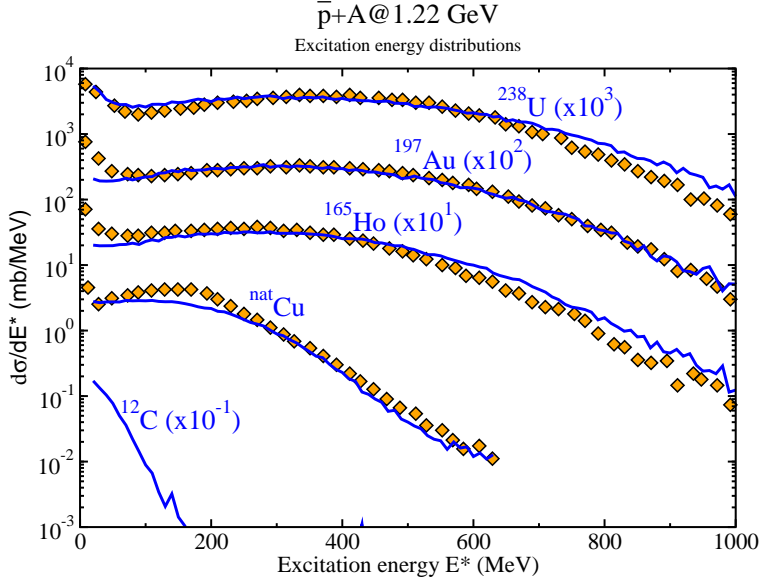}}}
\end{picture}
\caption{Inclusive excitation energy distributions for $\bar{p}$-induced 
reactions at an incident energy of 1.22 GeV on various targets, as indicated. 
The GiBUU+SMM calculations 
(solid curves) are compared to available experimental data (filled symbols), 
taken from~\cite{fradata}. The curves are multiplied by successive powers of 
$10$ starting from ${}^{\rm nat}$Cu.
}
\label{Fig2}
\end{figure}

With the improved initial target configurations reactions with antiproton 
beams have been performed, first, at low incident energies of $1.22$ GeV, 
where data on excitation energy and fragment yields are available~\cite{fradata}, 
and then at higher energies around the $\Xi\bar{\Xi}$ production 
threshold ($P_{\rm lab}=2.62$ GeV/c beam momentum or $E_{\rm lab}=1.84$ GeV incident 
energy). In the latter case complementary reactions with $\Xi$-beams have also performed. 
The properties of the residual source have been extracted by counting only bound 
particles, as explained in the previous section. After freeze-out we apply 
the SMM model~\cite{smm} to simulate the decay of the excited residual source. 
These calculations are denoted as GiBUU+SMM. 

\subsection{Fragment production in antiproton-induced reactions}

\begin{figure}[t]
\unitlength1cm
\begin{picture}(10.,8.0)
\put(1.5,0.){\makebox{\psfig{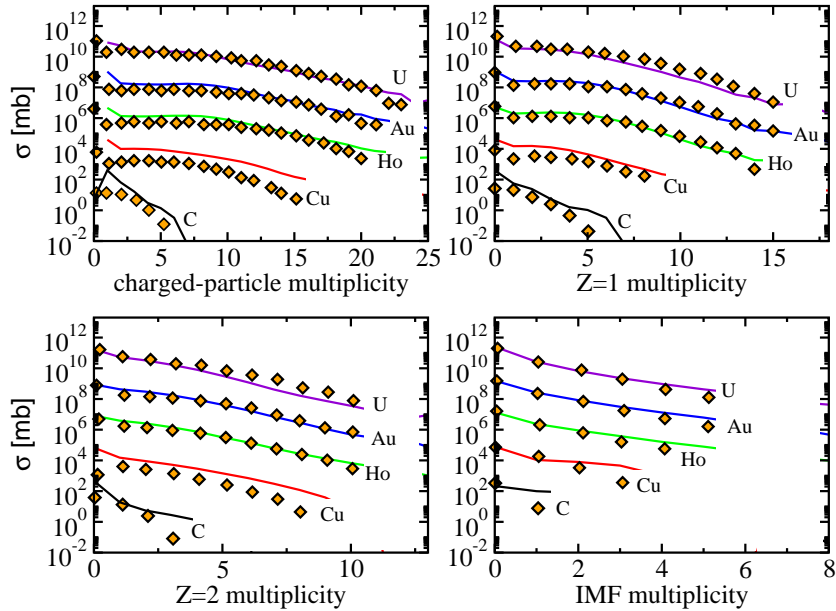}}}
\end{picture}
\caption{Inclusive fragment multiplicity distributions for $\bar{p}$-induced 
reactions at an incident energy of 1.22 GeV on different targets, as indicated.  
Upper panel on the left: charged-particles, upper panel on the right: 
protons and $Z=1$-fragments, lower panel on the left: $Z=2$-fragments, and 
lower panel on the right: intermediate mass fragments (IMF). The 
GiBUU+SMM calculations (solid curves) are compared with the data~\cite{fradata} 
(filled symbols). The curves are multiplied by successive powers of $10^{2}$ 
starting from $C$.
}
\label{Fig3}
\end{figure}

Fig.~\ref{Fig2} shows the theoretical results for the inclusive excitation 
energy distributions in comparison with experimental data. The shape and 
also the absolute values of the experimental distributions are reproduced 
fairly well by the transport calculations. 
The application of the SMM code after freeze-out provides single nucleons and 
fragments. Fig.~\ref{Fig3} shows the particle yields as the result of the 
combined GiBUU+SMM approach. The comparison with the experimental data for 
various types of fragments - in shape and absolute values - is very good, 
except for the very light systems. Note that the cases $Z=1$ and 
charged particles include also emitted protons from the pre-equilibrium 
GiBUU stage. 

Similar complementary fragmentation studies have been performed for 
proton-induced reactions and for spectator fragmentation in 
heavy-ion collisions~\cite{gaitanos2}, again with 
a satisfactory description of fragment yields and their spectra. In the 
following we will explore also the formation of hypernuclei. Therefore, apart from 
the requirement of a correct description of the fragmentation, a reliable dynamical 
description of particle production is also very important, in particular, for 
resonances as well as for pions, kaons and hyperons. 
Particle production in antiproton-induced reactions has been found to work well, 
as discussed in detail in Refs.~\cite{larionov,larionov2}. At this level 
of investigation we conclude that the combination of the GiBUU transport model 
together with the SMM approach represents a reliable method for the dynamical 
description of the entire process, starting from the pre-equilibrium dynamics 
up to the clusterization of the excited system. 

\subsection{Proton- versus antiproton-induced reactions}
 
\begin{figure}[t]
\unitlength1cm
\begin{picture}(10.,8.0)
\put(1.5,0.){\makebox{\psfig{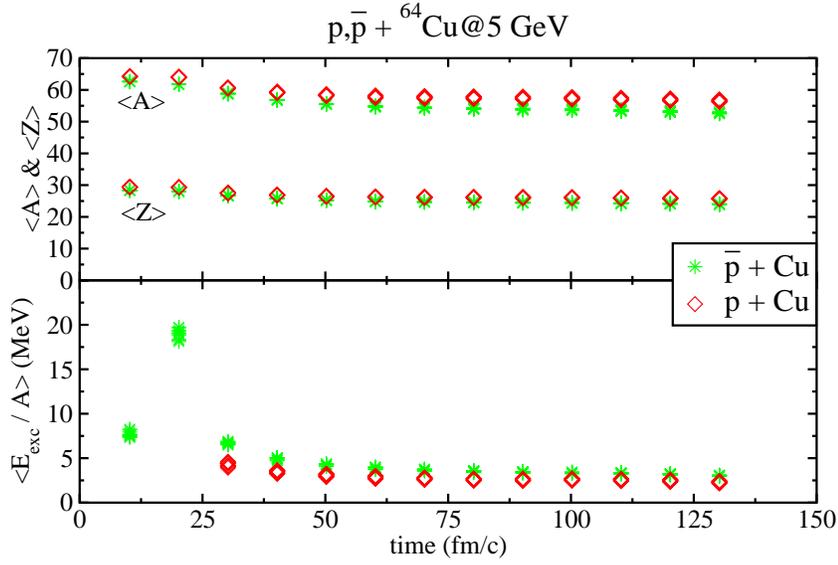}}}
\end{picture}
\caption{Time evolution of the average mass $\langle A \rangle$, the 
average charge $\langle Z \rangle$ (upper panel, as indicated) and 
the average excitation energy per nucleon 
$\langle E_{\rm exc}/A \rangle$ (lower panel), as 
the result of GiBUU calculations, 
for proton-induced (open diamonds) and antiproton-induced (filled stars) 
reactions at an incident energy of 5 GeV and impact parameter of b=3.4 fm. 
For the p+Cu reaction the excitation energy at early times 
is $\simeq 40$ MeV/nucleon and not visible in the lower figure.}
\label{Fig4}
\end{figure}
Reactions induced by protons~\cite{rudy} and heavy-ions~\cite{gaitanos1} 
turned out to be a useful tool for studies on single-$\Lambda$ hypernuclei. 
However, the formation of double-strange hyperfragments has not been 
investigated in detail, neither experimentally nor theoretically, so far. 
As proposed by the PANDA Collaboration, reactions with antiprotons 
are expected to be an ideal environment for the production of double-$\Lambda$ 
hypernuclei~\cite{panda1,panda2,panda3,ferro}. At first, by looking at typical 
properties such as 
mass and charge numbers and excitation of the bound system, no differences 
are visible between proton- and antiproton-induced reactions. This is 
demonstrated in Fig.~\ref{Fig4} in terms of the time evolution of these 
quantities at an incident energy of $E_{\rm lab}=5$ GeV. In the p-induced 
reactions the protons penetrate deeply into the nucleus, and the energy transfer 
is caused by multiple rescattering with resonance production and absorption 
of mesons. 

A quite different reaction scenario is found in antinucleon-induced reactions. 
The strong absorption confines interactions to a small layer at the nuclear surface. 
Hence, in the $\bar{p}+\rm Cu$ case the antiprotons do not penetrate much into  
the nuclear interior, and annihilate mainly close to the nuclear surface. 
In a $\bar{p}N$ annihilation final states with many mesons are locally 
produced~\cite{clover}. 
This creates a peak in the excitation energy at $t\sim 20$ fm/c, see 
Fig.~\ref{Fig4}, due to the excitation of the residual nucleus by annihilation mesons. 
Multiple meson-nucleon re-scattering is also possible in this case. Thus one 
expects a similar energy transfer as in the proton-nucleus case at long time 
scales, as also shown in Fig.~\ref{Fig4}.

The major difference between the proton-nucleus and antiproton-nucleus cases 
appears in the multiplicities of various produced particle species. 
While in the first case final channels with only few mesons 
(pions, kaons, antikaons) are mostly probable, the predominant channel in 
antiproton-nucleus 
collisions is the annihilation into many-body final states. 
Thus, not only mesons (mostly pions, and then kaons/antikaons), but also 
many-body final states with hyperons/antihyperons can be created in annihilation. 
\begin{figure}[t]
\unitlength1cm
\begin{picture}(10.,8.0)
\put(1.5,0.){\makebox{\psfig{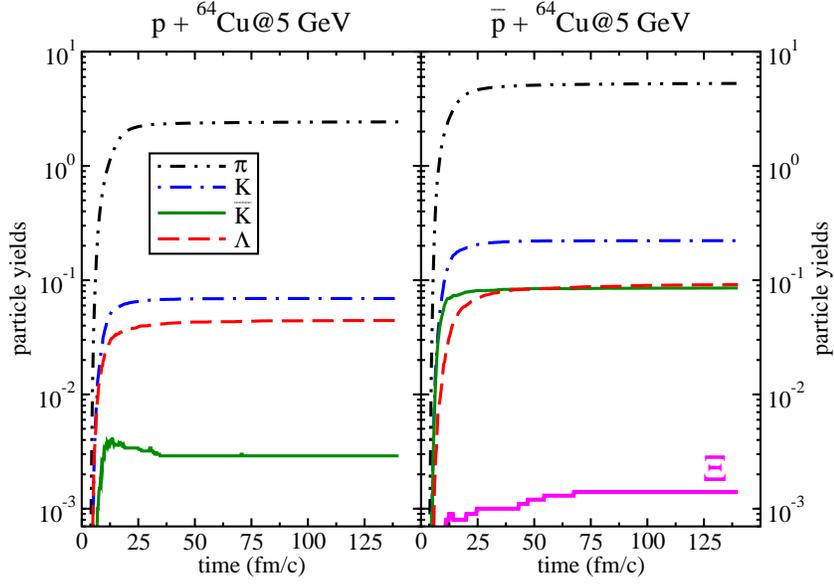}}}
\end{picture}
\caption{GiBUU calculations for the yields of various particle species 
(as indicated) as function of time in proton-induced (left panel) and 
antiproton-induced (right panel) reactions at an incident energy of 
5 GeV and impact parameter of b=3.4 fm. The cascade $\Xi$-particle, 
which is indicated inside the right panel, is produced only in 
antiproton-induced events. 
}
\label{Fig5}
\end{figure}
This difference in the particle yields between proton- and antiproton 
induced reactions can be seen in Fig.~\ref{Fig5}. Most of the particle yields 
are enhanced by almost a factor of two in the $\bar{p}$-induced reactions, 
in particular the increase of the antikaon yield is remarkable. In fact, 
in the $p$-nucleus 
case antikaons can be formed only in $4$-body final states in the process 
$pN\rightarrow NNK\bar{K}$ which has a very low branching ratio. 
On the other hand, in the $\bar{p}$-case 
the production of $K\bar{K}$-pairs in pure mesonic annihilation events 
is possible. Thus the 
$\bar{K}$-yield dramatically rises in antiproton-induced reactions. 
Another difference between the two reaction cases appears in the production of the 
cascade $\Xi$-particle, which appears only in antiproton-nucleus collisions. Its 
yield is extremely low because of the low production cross section for the 
primary process $\bar{p}p\rightarrow \Xi\bar{\Xi}$~\cite{cern1}. 
Note that cascade particles can be produced in secondary processes involving 
antikaons, mostly in the process $\bar{K}N\rightarrow \Xi K$~\cite{cern2}.

\subsection{Production of $S=-2$ hypernuclei in $\bar{p}+A$-reactions}

Our study shows that in antiproton-induced reactions a copious production of hyperons 
is possible, which enhances the probability for the production of hypernuclei, 
in particular, those with two captured $\Lambda$ hyperons. A particular candidate for 
the formation of double-strange matter is expected to be the cascade particle. However, 
an answer on the question, whether the cascade particles give the major contribution 
to the formation of double-strange hypernuclei or not, depends on two issues. 
First, the $\Xi$ particles have to be slow in order to be captured inside 
the residual system, otherwise the relativistic mean-field, in particular 
its Lorentz-vector component, is too repulsive and the resulting effective 
potential too shallow for a capture into a bound state. 
Secondly, the cross section for the process $\Xi N\rightarrow \Lambda\Lambda$ must 
be high enough, and, at least, comparable with the corresponding elastic 
$\Xi N\rightarrow \Xi N$ and also quasi-elastic $\Xi N\rightarrow \Lambda\Sigma$ processes.

\begin{figure}[t]
\unitlength1cm
\begin{picture}(10.,8.0)
\put(1.75,0.){\makebox{\psfig{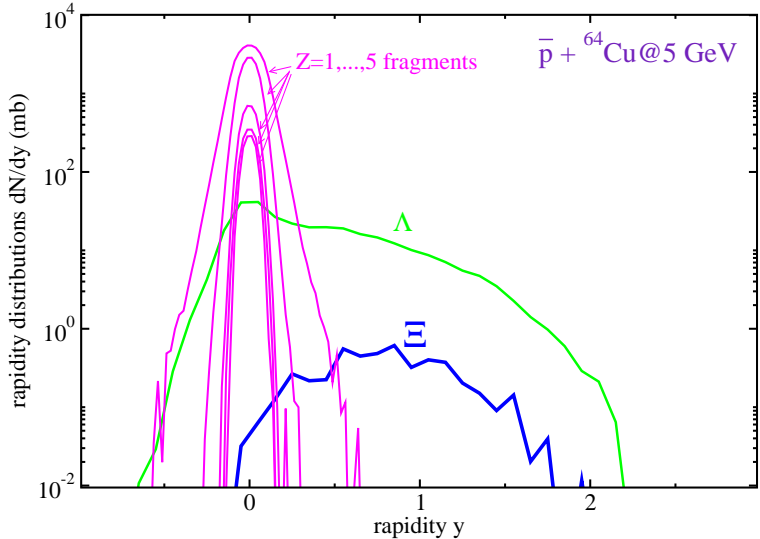}}}
\end{picture}
\caption{GiBUU+SMM calculations for the rapidity distributions of fragments 
with charge $Z=1,\cdots, 5$ and hyperons with strangeness $S=-1$ ($\Lambda$) and 
$S=-2$ ($\Xi$), as indicated, for inclusive $\bar{p}+\rm Cu@5$~GeV reactions.
}
\label{Fig6}
\end{figure}

The PANDA Collaboration~\cite{panda1,panda2,panda3} has proposed an experiment 
consisting of two targets. The primary target serves for the production of the 
cascade particles through the interaction of the antiproton beam with the target 
nucleons. The emitted $\Xi$-particles will be decelerated by ionization energy 
loss in ordinary matter before their interaction with a secondary target. 
In this way one assumes that the slow moving $\Xi$-particles will 
be captured by the secondary target, and the formation of double-strange 
hypernuclei will occur after the conversion of the bound $\Xi$'s into two 
$\Lambda$-hyperons inside the secondary residual nucleus by the collision with 
a nucleon. 

Before analyzing the scenario with the secondary target, i.e., performing 
GiBUU studies for $\Xi$-induced reactions, it is useful to investigate 
the strangeness dynamics, in particular the properties of the produced 
$\Xi$ hyperons, from the antiproton interaction with a primary target. We 
have performed GiBUU+SMM calculations for inclusive $\bar{p}$-interactions with 
a ${}^{64}$Cu target nucleus at an incident energy of $E_{lab}=5$ GeV. 

Fig.~\ref{Fig6} shows the rapidity spectra of various fragments 
and single hadrons produced in $\bar{p}+Cu$ reactions at an incident energy 
of $E_{\rm lab}=5$ GeV. Various intermediate mass fragments (and also heavier 
residues not shown in this figure) are produced close to the rest frame of 
the target. The rapidity distributions of various particles with strangeness 
are quite different. Hyperons with $S=-1$ ($\Lambda$ particles) show a broad 
spectrum in rapidity, while particles with $S=-2$ ($\Xi$ particles) are rather 
located at high rapidities. The difference between the rapidity 
spectra of hyperons with $S=-1$ and $S=-2$ arises from the different 
way how these particles are produced. As discussed above, $\Lambda$ particles 
can be produced in various processes. Direct production in the annihilation 
channel $\bar{p}p\rightarrow\bar{\Lambda}\Lambda (M)$ in first 
collisions provides the high energy tail of the rapidity distribution. 
However, $\Lambda$ particles undergo multiple secondary scattering with nucleons 
which redistributes their velocities. On the other hand, 
the channel $\bar{K}\Lambda\rightarrow \Xi N$ 
gives the major contribution to $\Xi$ production. Such type of a secondary process 
is possible only when the energy of the initial antikaon is high, because 
of the high threshold $\sqrt{s_{\rm thr}}=M_{\rm \Xi}+M_{\rm N}=2.253$ GeV 
requiring a minimal kinetic energy of $E_{\rm kin}\simeq 2.13 (1.7)$ GeV for the antikaon 
($\Lambda$ hyperon) in the cm-frame. 
Thus, the $\Xi$ production is shifted to high rapidities. 
The produced cascade particles can leave the bound system, while most of the $\Lambda$ 
hyperons are captured. In addition to collisions, the relativistic 
mean-field plays also an important role for the capture of hyperons, 
in particular, for the low energetic ones, as discussed in more detail 
in Ref.~\cite{larionov2}.

\begin{figure}[t]
\unitlength1cm
\begin{picture}(10.,8.0)
\put(1.75,0.){\makebox{\psfig{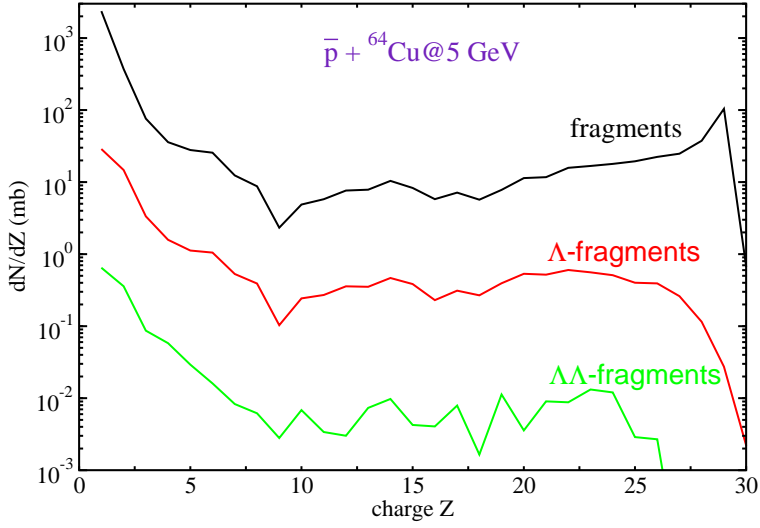}}}
\end{picture}
\caption{GiBUU+SMM calculations for the charge distributions of 
nuclear fragments (upper curve), of single-$\Lambda$ 
(middle curve) and double-$\Lambda$ (lower curve) hyperfragments.
}
\label{Fig7}
\end{figure}
Fig.~\ref{Fig7} shows the results for the charge distributions of nuclear 
fragments, single-$\Lambda$ and also double-$\Lambda$ hypernuclei, as 
indicated. The nuclear fragments show the typical distribution with the 
evaporation peak close to $Z\simeq Z_{\rm targ}$, the fission hill around 
$Z\simeq Z_{\rm targ}/2$ and the power-low ($Z^{-\tau}$) multifragmentation 
region at smaller Z ($2<Z<8$). The formation of hyperclusters arises 
mainly due to the momentum coalescence of the captured $\Lambda$-particles 
with the SMM-fragments. However, the yield of light double-strange 
hyperfragments is around three orders of magnitudes less relative to that of 
nuclear fragments. We note that the inclusion of the mean-field for 
the hyperons is important for the production of hypernuclei. Indeed, 
$\Lambda$-hyperons feels an attractive in-medium optical potential at low energies, 
which at saturation density is around $-38$ MeV. This attractive potential is 
also responsible for the binding mechanism of the slow-moving $\Lambda$-particles 
inside the residual target nucleus, as discussed in detail in Ref.~\cite{larionov2}.

\begin{figure}[t]
\unitlength1cm
\begin{picture}(10.,8.0)
\put(1.75,0.){\makebox{\psfig{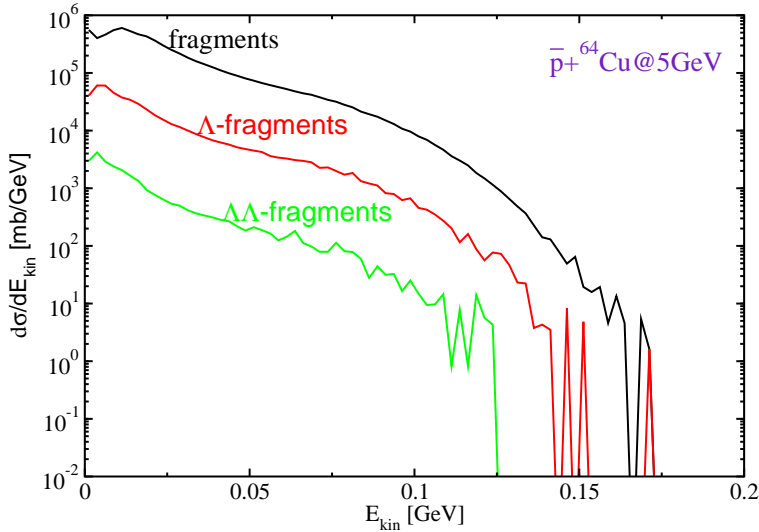}}}
\end{picture}
\caption{Same as in the previous Fig.~\ref{Fig7}, but now for the 
kinetic energy spectra.
}
\label{Fig8}
\end{figure}

Fig.~\ref{Fig8} shows the kinetic energy distributions for nuclear fragments and 
hyperfragments with $S=-1$ and $S=-2$ strangeness content. Again the yield of 
double-$\Lambda$ hypernuclei is very low, as already shown in the previous 
figure. Fig.~\ref{Fig8} demonstrates that various particles are emitted from a 
thermalized residual source, since the slopes of various particle types are 
very similar. 

\subsection{Secondary hypernuclear formation by $\Xi$-beams}

So far we have considered the properties of hypermatter in collisions 
between antiprotons with only the first target. As an intermediate result, 
a coalescence between fragments of the primary target with cascade particles 
occurs, however, with very low probability due to the high velocities of the 
produced cascade particles. In the considered example $\bar{p}+\rm Cu@5$ GeV 
the maximum of the rapidity distribution (see again 
Fig.~\ref{Fig6}) of the $\Xi$-particles lies around $y\simeq 1$ corresponding 
to a kinetic energy (momentum) of $E_{\rm kin}\simeq 0.7$ GeV ($p\simeq 1.54$ GeV/c). 

\begin{figure}[t]
\unitlength1cm
\begin{picture}(10.,8.0)
\put(1.5,0.){\makebox{\psfig{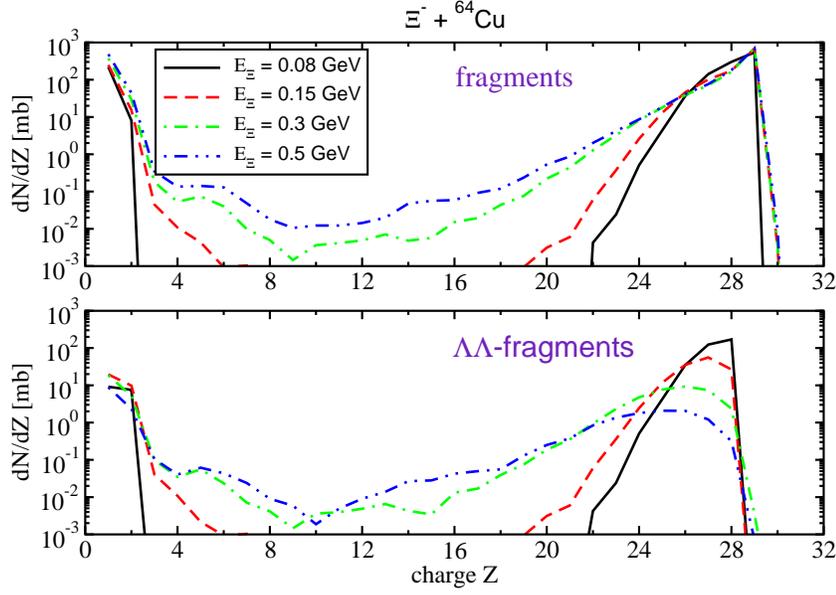}}}
\end{picture}
\caption{GiBUU+SMM calculations for the charge distributions of nuclear 
fragments (upper panel) and double-$\Lambda$ clusters (lower panel)
for $\Xi^{-}+{}^{64}$Cu reactions at various kinetic energies of the 
$\Xi$-beam in the laboratory frame, as indicated.
}
\label{Fig9}
\end{figure}

In order to estimate the dynamical properties of the collisions with the 
secondary target, we have performed complementary transport calculations 
with $\Xi^{-}$-beams on the same target (${}^{64}$Cu) at incident kinetic 
energies of $E_{\rm lab}=0.08,0.15,0.3,0.5$ GeV which correspond to the region just 
below the maximum value of the rapidity of the $\Xi$-particles.  
Fig.~\ref{Fig9} shows the results for $\Xi^{-}+\rm Cu$-reactions at the beam energies 
as indicated. The nuclear fragment charge distributions reveal an U-shape typical 
for the quasi-fission and quasi-evaporation decay modes of medium heavy nuclei 
(A$\sim ~60-100$)~\cite{smm} at low excitation energies, in contrast to the 
multifragmentation pattern of Fig.~\ref{Fig7}. Another important difference with 
respect to the $\bar{p}$-induced reactions is that now each reaction event 
contains a $\Xi$ particle, which collides with the bound nucleons producing 
two $\Lambda$ particles and, therefore, leading to the production of 
$\Lambda\Lambda$-hyperfragments with large probability. Thus, the 
distributions of the nuclear fragments and of the double-$\Lambda$ hypermatter 
are very similar in the absolute values, as seen in Fig.~\ref{Fig9}.
As a general trend, the production of double-strange hypernuclei decreases 
with increasing $\Xi$ energy (see again lower panel in Fig.~\ref{Fig9}). 
In particular, the abundance of heavy hyperfragments reduces significantly 
as the beam energy rises. 
Since heavy residues mostly survive in peripheral events, the probability to form 
a $\Lambda\Lambda$-hypernuclei obviously strongly decreases with increasing 
$\Xi$-beam momentum.

\begin{figure}[t]
\unitlength1cm
\begin{picture}(10.,8.0)
\put(1.75,0.){\makebox{\psfig{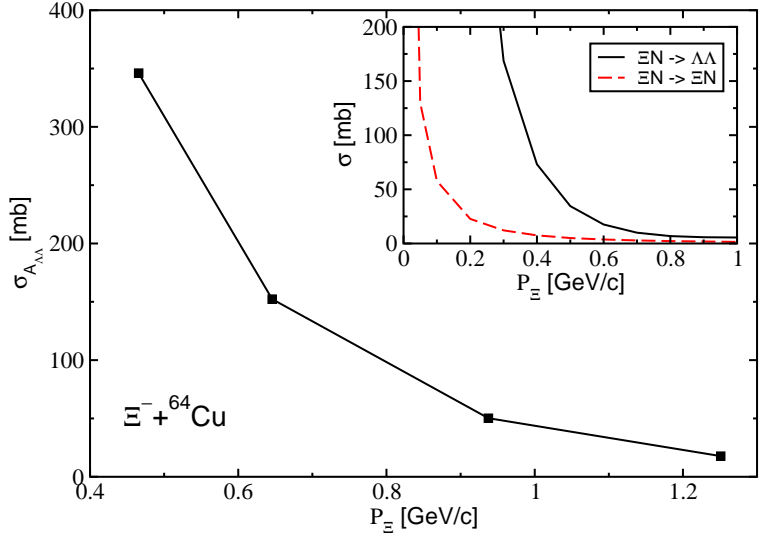}}}
\end{picture}
\caption{(Main panel) Production cross section for double-strange 
hypermatter as function of the $\Xi$-beam momentum. The insert panel 
shows the elementary cross sections~\cite{esc04} for the indicated 
elementary channels.
}
\label{Fig10}
\end{figure}

In total, it turns out that the double-strange hypernuclei yield 
produced in reactions between the $\Xi$-beam and the secondary target 
strongly decreases with increasing beam energy (see Fig.~\ref{Fig10}). 
This is because of the quickly dropping 
$\Xi^{-} p\rightarrow \Lambda\Lambda$ elementary cross section as compared 
to the elastic one. This is shown in the inner panel in Fig.~\ref{Fig10} for 
theoretical calculations according to the extended soft core 
model~\cite{esc04} for the elastic $\Xi N\rightarrow \Xi N$ and inelastic 
$\Xi N\rightarrow \Lambda\Lambda$ channels. Indeed, for momenta higher 
than $0.4$ GeV/c the relevant inelastic channel strongly drops and becomes 
comparable with the elastic cross section. 
We conclude that in order to measure double-strange hypernuclei with 
very high cross section, the cascade particles produced in collisions 
between the antiproton beam with the primary target should have as low 
energy as possible. Indeed, in the PANDA experiment the emitted cascade 
particles are supposed to traverse a material for deceleration, before 
the collision with the secondary target system.

\begin{figure}[t]
\unitlength1cm
\begin{picture}(10.,8.0)
\put(1.75,0.){\makebox{\psfig{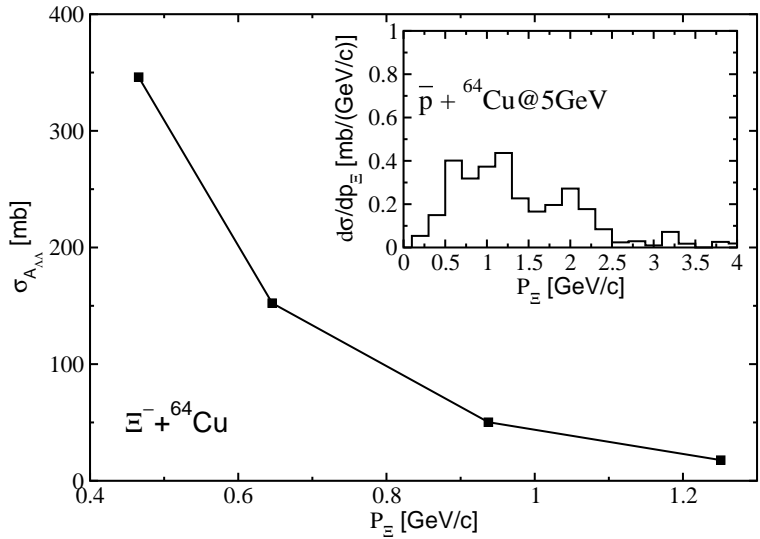}}}
\end{picture}
\caption{(Main panel) Same as in Fig.~\ref{Fig10}. The insert panel 
shows now the $\Xi$-production cross section from $\bar{p}$-collisions 
on the first target, as indicated. 
}
\label{Fig11}
\end{figure}

For the determination of the total production cross section of double-strange 
hypernuclei one needs the momentum distribution of the $\Xi$-particles emitted 
from the antiproton-reaction with the primary target. Fig.~\ref{Fig11} shows 
this in the insert panel, again together with the double-strange hypernuclear 
cross section. As one can see from Fig.~\ref{Fig11}, the region 
around $P_{\rm \Xi} \simeq 0.5-1.2$ GeV/c will contribute mostly to the total production 
cross section, for the considered $\bar{p}+{}^{64}\rm Cu@5$ GeV reaction. Thus, the 
effective value of the total double-strange hyperfragment cross section is estimated 
to be of the order of several mb, as the result of a momentum integration 
folding between $\sigma_{A_{\Lambda\Lambda}}(p_{\Xi})$ and 
$\frac{d\sigma}{dp_{\Xi}}$ and after consideration of attenuation of the 
$\Xi$ hyperon due to its finite lifetime.

\section{Conclusions and outlook}
\label{sec4}

In summary, we have studied the formation of hypernuclei in reactions 
relevant for the PANDA project at the new FAIR facility. We have investigated 
hypernuclear production in two steps, as suggested in the proposal 
of the PANDA experiment, using the GiBUU transport theoretical 
framework for the description of the pre-equilibrium dynamics, and 
the SMM model for fragment formation. A supplementary momentum 
coalescence between hyperons and SMM-fragments has been adopted for 
the production of hyperfragments. For antinucleon-nucleon 
interactions an extension of the GiBUU approach was necessary by 
including more binary processes involving the scattering 
between nucleons and hyperons.

The combined GiBUU+SMM approach had earlier been shown to work well concerning 
the spectra and yields of produced strange particles and nuclear fragments 
in low energy antiproton-nucleus reactions. 
We have applied then the same model to antiproton induced reactions at higher 
energies. It was found that a simple 
coalescence between nuclear clusters and $\Lambda$-hyperons to single-strange 
hypernuclei is possible already in antiproton collisions with a first target. 
However, the formation of double-strange hypernuclei and the capture of 
double strange $\Xi$-particles occurs with very low probability, due to 
the high energies of the produced cascade hyperons. Thus, as suggested 
by the PANDA experiment, the $\Xi$ particles have to be decelerated in 
order to be captured into a secondary target. This scenario is supported 
by the present transport simulations. The main mechanisms leading to the 
formation of double-$\Lambda$ hypernuclei are the attractive mean-field 
and, in particular, the inelastic process $\Xi N\rightarrow \Lambda\Lambda$, 
whose cross section dramatically rises with decreasing $\Xi$ energy. 

At this level of investigation the predictive power of the theoretical calculations 
on hypernuclei is to some extend limited by incomplete knowledge on $YN$- and 
$YY$-interactions in free space and in nuclear matter. 
However, we are confident about our results  by recalling that the underlying 
processes of strangeness production and fragmentation are well described, reproducing 
existing data very satisfactorily. 
We have estimated the production cross sections of double strange hypernuclei as 
function of the $\Xi$-hyperon energy by its capture with the secondary target. As an 
important result, double-strange hypernuclear production at PANDA is possible with 
high probability, if low energy cascade-particle beams will be used. We emphasize 
the relevance of our theoretical results for the future activities at FAIR.

{\it Acknowledgments.} We would like to thank A. Botvina for providing us the code 
for the Statistical Multifragmentation Model (SMM). This work is supported 
by BMBF contract GILENS 06, DFG contract Le439/9-1 and HIC for FAIR.


\end{document}